\documentclass[reprint,aps,prd,superscriptaddress,showpacs,preprintnumbers,amsmath,amssymb,floatfix]{revtex4-1} 
\pdfoutput=1

\usepackage[inline, shortlabels]{enumitem}
\setlist{
  listparindent=\parindent,
  parsep=0pt,
}
\usepackage[mathlines]{lineno}
\usepackage{graphicx,subfigure}
\usepackage{epstopdf}
\usepackage{dcolumn}
\usepackage{bm}
\usepackage{color}
\usepackage{upgreek}
\usepackage{afterpage}
\usepackage{tikz, pgfplots}
\usepackage{bigstrut}
\usepackage[colorlinks=true]{hyperref}
\usepackage[normalem]{ulem}

\newcommand{\gev}{\nolinebreak GeV/$c^2$}
\newcommand{\keVnrNS}{keV} 

\usepackage{moreverb} 

\immediate\write18{texcount -inc -incbib
-sum SuperCDMS_HT.tex > wordcount.tex}

\immediate\write18{texcount -char -freq
 SuperCDMS_HT.tex > charcount.tex}

\renewcommand{\d}[1]{\ensuremath{\operatorname{d}\!{#1}}}

\def\ba{$^{133}$Ba}
\def\cf{$^{252}$Cf}
\def\pbsix{$^{206}$Pb}
\def\pbten{$^{210}$Pb}
\def\po{$^{210}$Po}

\newcolumntype{1}[1]{D{,}{\pm}{#1}}
\newcolumntype{2}[1]{D{,}{.}{#1}}
\newcolumntype{3}[1]{D{,}{-}{#1}}

\newcommand*\patchAmsMathEnvironmentForLineno[1]{%
   \expandafter\let\csname old#1\expandafter\endcsname\csname #1\endcsname
   \expandafter\let\csname oldend#1\expandafter\endcsname\csname end#1\endcsname
   \renewenvironment{#1}%
      {\linenomath\csname old#1\endcsname}%
      {\csname oldend#1\endcsname\endlinenomath}}%
\newcommand*\patchBothAmsMathEnvironmentsForLineno[1]{%
   \patchAmsMathEnvironmentForLineno{#1}%
   \patchAmsMathEnvironmentForLineno{#1*}}%
\AtBeginDocument{%
\patchBothAmsMathEnvironmentsForLineno{equation}%
\patchBothAmsMathEnvironmentsForLineno{align}%
\patchBothAmsMathEnvironmentsForLineno{flalign}%
\patchBothAmsMathEnvironmentsForLineno{alignat}%
\patchBothAmsMathEnvironmentsForLineno{gather}%
\patchBothAmsMathEnvironmentsForLineno{multline}%
}

\begin{document}


\title{Results from the Super Cryogenic Dark Matter Search (SuperCDMS) experiment at Soudan}

\author{R.~Agnese} \affiliation{Department of Physics, University of Florida, Gainesville, FL 32611, USA}
\author{T.~Aramaki} \affiliation{SLAC National Accelerator Laboratory/Kavli Institute for Particle Astrophysics and Cosmology, Menlo Park, CA 94025, USA}
\author{I.J.~Arnquist} \affiliation{Pacific Northwest National Laboratory, Richland, WA 99352, USA}
\author{W.~Baker} \affiliation{Department of Physics and Astronomy, and the Mitchell Institute for Fundamental Physics and Astronomy, Texas A\&M University, College Station, TX 77843, USA}
\author{D.~Balakishiyeva} \affiliation{Department of Physics, Southern Methodist University, Dallas, TX 75275, USA}
\author{S.~Banik} \affiliation{School of Physical Sciences, National Institute of Science Education and Research, HBNI, Jatni - 752050, India}
\author{D.~Barker} \affiliation{School of Physics \& Astronomy, University of Minnesota, Minneapolis, MN 55455, USA}
\author{R.~Basu~Thakur} \affiliation{Fermi National Accelerator Laboratory, Batavia, IL 60510, USA}\affiliation{Department of Physics, University of Illinois at Urbana-Champaign, Urbana, IL 61801, USA}
\author{D.A.~Bauer} \affiliation{Fermi National Accelerator Laboratory, Batavia, IL 60510, USA}
\author{T.~Binder} \affiliation{Department of Physics, University of South Dakota, Vermillion, SD 57069, USA}
% ...

\smallskip
\date{\today}

\collaboration{SuperCDMS Collaboration}

\noaffiliation


\smallskip

\begin{abstract}
We report the result of a blinded search for Weakly Interacting Massive Particles (WIMPs) using the majority of the SuperCDMS Soudan dataset. With an exposure of 1690 kg days, a single candidate event is observed, consistent with expected backgrounds. This analysis (combined with previous Ge results) sets an upper limit on the spin-independent WIMP--nucleon cross section of $1.4 \times 10^{-44}$~($1.0 \times 10^{-44}$)~cm$^2$~at 46~\gev. These results set the strongest limits for WIMP--germanium-nucleus interactions for masses $>$12~\gev.
\end{abstract}


\maketitle


Astrophysical observations indicate that the matter content of the universe is dominated by non-baryonic, cold dark matter (DM)~\cite{Plank2015}. Weakly Interacting Massive Particles (WIMPs) are a favored class of dark matter candidates~\cite{Goodman1985}, and their thermal production in the early universe would yield a relic density that is consistent with the observed matter abundance. The weak interaction of WIMPs with normal matter would
enable their detection in laboratory experiments~\cite{Goodman1985} via elastic scattering with nuclei,
yielding an approximately exponential energy spectrum~\cite{Lewin1996}.

We present the results of a search for DM scatters off atomic nuclei using 15 interleaved Z-sensitive Ionization- and Phonon-mediated (iZIP) detectors~\cite{r133izip} of the SuperCDMS Soudan experiment.  It employs the Cryogenic
Dark Matter Search (CDMS II)~\cite{Ahmed2010} low-background apparatus~\cite{Akerib2005}, which consists of a cryostat surrounded by a passive shield and outer muon veto situated beneath an overburden of 2090 meters water equivalent.  The passive shield comprises 40\,cm of outer polyethylene, 22.5 cm of lead, and 10 cm of inner polyethylene.  The cryostat and internal cold hardware provide an additional 3 cm of copper shielding. Each 0.6 kg iZIP detector consists of a 76-mm diameter, 25-mm thick, cylindrical, high-purity germanium substrate in which a recoiling nucleus or electron creates electron-hole pairs and phonons.  An applied electric field
(bias), parallel to the cylindrical axis in the bulk and transverse to that axis near the faces, causes electrons and holes to drift to inner disklike and outer annular electrodes on the two faces.  Four phonon sensors are distributed on each face.

For each event, we reconstruct two energy parameters: (1)~``ionization energy,'' which is the number of electron-hole pairs collected, converted to energy units, and is estimated from the combination of electron and hole information, and (2)~``recoil energy,'' which is obtained by subtracting from the total phonon energy an ionization-signal-derived estimate of Neganov-Trofimov-Luke phonon energy~\cite{Luke1988, Neganov1985}.  The ratio of ionization energy to recoil energy is ``ionization yield.''  Because it is suppressed for nuclear recoils  relative to electron recoils by a factor of $\approx$3 in germanium, ionization yield is the key parameter discriminating nuclear recoils (e.g., due to dark matter) from background-induced electron recoils.

Because we may misidentify electron recoils with suppressed ionization collection as nuclear recoils, we exclude regions near the surface of the detector for which ionization collection is incomplete using four radial- and $z$-position proxies: (1)~``ionization radial partition,'' the ionization signal in the outer electrode divided by the sum of the outer and inner electrode signals, with one estimate each from the hole and electron collection faces, (2)~a ``phonon radial partition'' constructed in an analogous fashion, (3)~``ionization $z$ partition,'' the difference in electron and hole ionization energy estimates divided by their optimal combination, and (4)~a ``phonon $z$ partition'' analogue.

A combination of the event parameters defines a ``fiducial volume'' inside each detector, within which we search for nuclear recoils.  Events inside the fiducial volume are labeled ``bulk'' while those outside are labeled ``surface.''  This procedure is termed ``fiducialization'' hereafter.  An early, conservative fiducialization study~\cite{r133izip} yielded a very low probability for misidentifying surface electron recoils as bulk nuclear recoils: $<$$1.7\,\times\,10^{-5}$ for 8--115 \keVnrNS. The excellent iZIP background rejection allows for a nearly background-free search, which makes effective use of a given experimental exposure while being robust to background systematics.  To maximize sensitivity, we optimize the fiducial volume, trading off between signal acceptance and expected misidentified background.  An analysis with an 8~keV threshold is most sensitive to DM masses $>$10~\gev.

We use datasets taken from March 2012 through July 2014.   Approximately 70\% of this time was used for DM-search data, while 10\% was used for calibration and the remaining 20\% was lost to experimental maintenance and periods of high detector noise. The total raw live time is 534 d. We removed data in which detectors were not functioning normally, yielding a total exposure of $1690$ kg days. Data taken multiple times per week with a \ba{} gamma-ray source provides a high-statistics electron-recoil sample for estimating background misidentification. The ionization and phonon energies were calibrated using the \ba{} 356 keV line and checked with the 10.36~keV Ge-activation line, which was recovered to $\approx$5\% accuracy~\cite{welliver2016}.  Every few months, we took data with a \cf{} neutron source to produce a sample of nuclear recoils to measure signal acceptance.

The region of parameter space used to search for nuclear recoils from DM interactions is defined using neutron calibration data.  In order to minimize bias, we excluded (``blinded'') this region prior to defining signal-acceptance and background-rejection criteria. An event was blinded if its energy exceeded a time-varying threshold value (3$\sigma$ above the mean of the noise distribution), its recoil energy was below 150~\keVnrNS, it was not identified as being due to low-frequency noise or an electronics glitch, its ionization partition parameters placed it within a loosely defined fiducial volume, it deposited energy in only a single detector, and its ionization yield lay within a loosely defined nuclear-recoil acceptance region.  Specific time periods not used for the DM search were left completely unblinded for special studies, as were the two to three days following neutron calibrations due to elevated backgrounds from a germanium electron-capture peak at 10.36~keV~\cite{Hampel1985}. Only data that remained blinded throughout our analysis was eligible for inclusion in the signal dataset; we excluded data considered in the prior, low-mass analysis~\cite{r133lowthreshold} because it was no longer blinded.

\begin{figure}[t!!]
\begin{center}
\includegraphics[width=250pt]{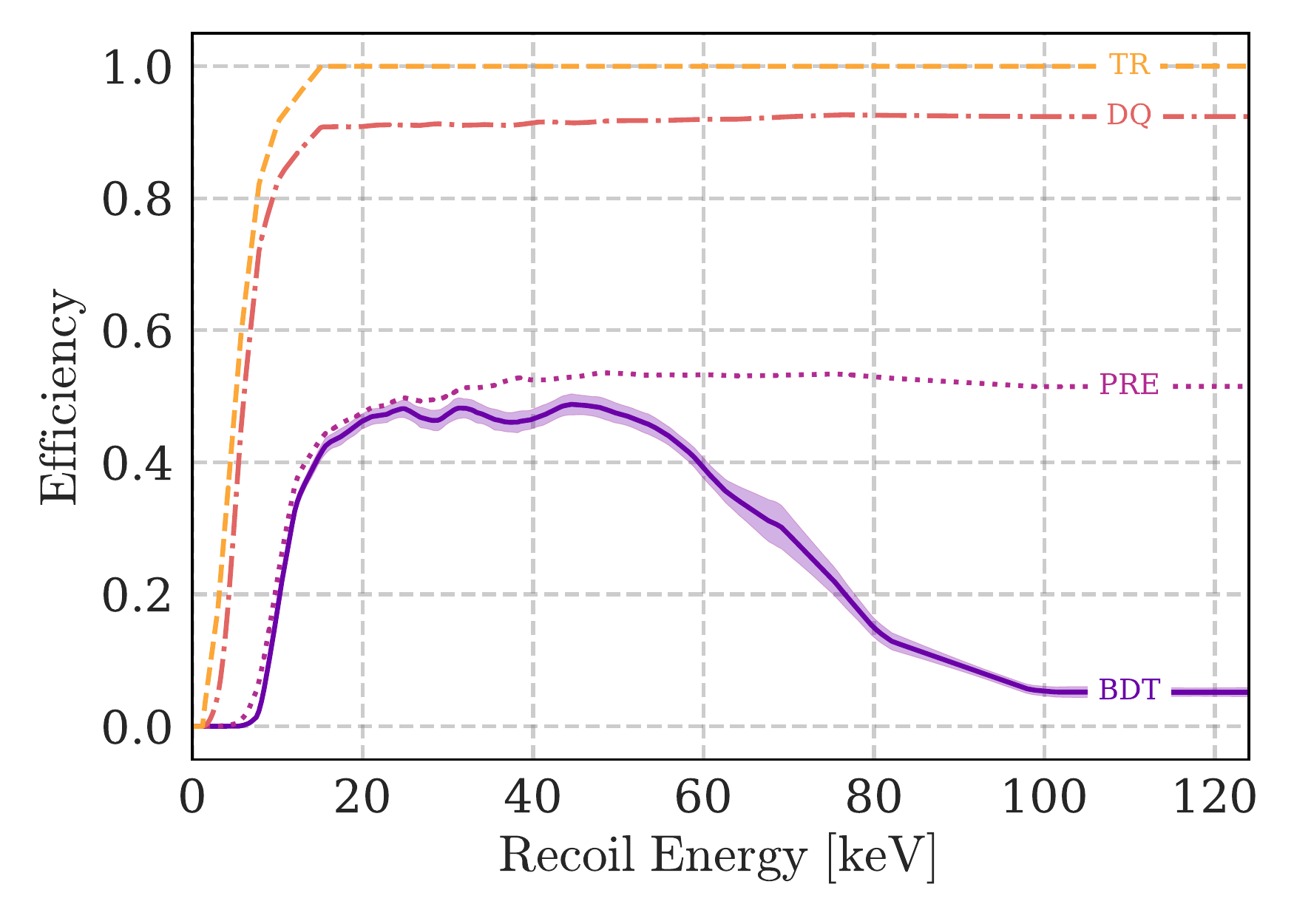}
\caption{The total exposure-weighted efficiency is shown after sequential application of event selection criteria, averaged over all detectors. From top to bottom: hardware phonon trigger (TR), data quality (DQ), event preselection (PRE), and BDT discrimination (BDT).  A 68\% CL uncertainty band on the overall efficiency is shown. }
\label{fig:efficiency_exposure}
\end{center}
\end{figure}

Figure~\ref{fig:efficiency_exposure} shows the hardware phonon \textit{trigger efficiency} as a function of energy, measured using the fraction of multiple-scatter \ba{} events in a detector that also triggered in the detector.

\textit{Data quality} cuts exclude events with erroneous or unreliable reconstructions from further analysis.  For every ionization and phonon signal, we calculate an energy- and time-dependent goodness-of-fit statistic for three hypotheses --- interaction event, low-frequency noise, and electronics glitch --- allowing removal of events inconsistent with particle interactions.  Figure~\ref{fig:efficiency_exposure} shows the efficiency of these data quality cuts.

We define a set of \textit{preselection} cuts by excluding events inconsistent with  aspects of the DM scattering hypothesis.  The rate of DM  multiple scatters would be negligibly small, so we discard events in which multiple detectors showed energy deposits with  $>$3$\sigma$ inconsistency with their respective noise distributions.
For each detector, the analysis threshold is the largest of the 95\% trigger efficiency energy, the blinding lower energy limit, or a fixed value of 4 \keVnrNS. We also reject any event coincident with activity in the muon veto due to the potential for nuclear recoils of muon-created particles.

Another set of preselection cuts provides loose fiducialization.  A set of loose cuts in ionization partitions remove the majority of surface events.  We also use a one-class support vector machine~\cite{scikit-learn, Scholkopf2001} to reject the 1\% of events least consistent with the neutron-calibration population in phonon partitions.  We require that the ionization signal be $>$$5\sigma$ above the mean of the noise distribution constructed from random trigger events.
We also observed in \ba{} calibration data a set of events suppressed in ionization yield, localized in phonon partition, uniformly distributed in time, and present in all detectors.  Cuts in phonon partition exclude this class of events, with a 15\% loss of fiducial volume.

The final preselection cut defines an energy-dependent region in ionization yield consistent with nuclear recoils. The ionization yield distribution of the \cf{} data is fit to an energy-dependent Gaussian with center $y_\text{NR}(E)$ and width $\sigma_\text{NR}(E)$.
Events within $3\,\sigma_\text{NR}(E)$ of $y_\text{NR}(E)$, specific to each detector and period, are retained.  Figure~\ref{fig:efficiency_exposure} illustrates the combined signal efficiency of the preselection cuts.

To identify DM candidates in the preselected dataset, an acceptance region is defined.  It is chosen by optimizing the sensitivity to the DM-nucleon spin-independent cross section given the expected signal characteristics and the backgrounds that might be misidentified as signal. We consider three background sources.

The first background is due to the broad continuum of Compton-scatter electron recoils (up to 2.5~MeV) produced by the gamma-ray background arising from natural radioactivity in our apparatus.  As mentioned earlier, the events most likely to be misidentified as nuclear recoils on the basis of ionization yield are those occurring in regions of the detector with incomplete ionization collection.  Ionization partition identifies events in these regions.

The second background arises from \pbten{} and its daughters.  Radon exposure during detector production and testing results in plate-out of \pbten{} on copper housings and detector surfaces.  During the multi-step decay of \pbten{} to the stable \pbsix{}, a variety of betas, x-rays, a 46.5 keV gamma-ray, and a 103~keV \pbsix{} daughter are emitted, yielding recoils near the detector surfaces.

The third background consists of neutrons.  Radiogenic neutrons arise from spontaneous fission and $(\alpha,n)$ reactions in our apparatus.  Cosmogenic neutrons arise from cosmic-ray muon spallation.  Not all of the latter can be rejected by the muon veto, as the parent muon may not pass through the muon veto panels.  Discrimination between neutron backgrounds and DM interactions is difficult given their similar (but not identical) energy spectra and spatial distributions in the detectors.

We perform the optimization using models for signal and backgrounds to determine acceptance and background misidentification as a function of cut values.  Reweighting our calibration datasets yields what we term ``model datasets.''

To build the signal model dataset, we first assign a weight to each \cf{} calibration event so that the spectrum of the reweighted data matches the shape of the theoretical DM recoil-energy spectrum for a particular DM mass~\cite{Lewin1996}, corrected by the energy-dependent efficiency of all cuts applied to this point.  We normalize these weights so their sum matches the spectrum-averaged exposure (SAE), defined as follows:
\begin{align}
SAE = M\,T \left. \int_{E_{\text{min}}}^{E_{\text{max}}} dE\, \epsilon(E)\, \frac{\d{R}}{\d{E}} \middle/ \int_{E_{\text{min}}}^{E_{\text{max}}} dE\, \frac{\d{R}}{\d{E}}\right.
\end{align}
where $M\,T$ is the experiment's raw exposure, $\epsilon(E)$ is the energy-dependent analysis efficiency, and $\frac{dR}{dE}$ is the expected DM differential recoil spectrum, evaluated for DM masses of 10, 25, 50, 100 and 250 \gev.

We use \ba{} calibration data to model the Compton-scatter background in the DM-search data, selecting events inside the $3\sigma_\text{NR}$~nuclear-recoil acceptance region as representative of Compton scatters with incomplete ionization collection (low ionization yield).  We construct weights by considering the single-scatter events in the \ba{} calibration data and the DM-search data within an ionization yield ``sideband,'' consisting of the region in ionization yield between the upper edge of the nuclear-recoil acceptance region and the lower edge of the full-collection electron-recoil band (defined using \ba{} data as a 3$\sigma$ band in similar fashion to the nuclear-recoil acceptance region definition).
We find weight functions of recoil energy and ionization radial and $z$ partition that, when applied to the \ba{} ionization-yield sideband dataset, produce distribution functions in these three parameters matching in shape to those of the DM-search ionization-yield sideband dataset.  We apply these weight functions to the single-scatter \ba{} events in the nuclear-recoil acceptance region to obtain the model dataset for this background.  In doing so, we assume that the weight functions of recoil energy and ionization partitions are independent of ionization yield.  Finally, we normalize these weights such that the sum of the weights of all events in this model dataset equals the expected number of single-scatter, DM-search, Compton-scatter events in the nuclear-recoil acceptance region, before any fiducialization, $N_\text{DM}^\text{NR}$.  This number is determined from $N_\text{Ba}^\text{NR}$, the number of single-scatter \ba{} events in the nuclear-recoil acceptance region; $N_\text{Ba}^\text{SB}$, the number of  single-scatter, \ba{} events in the ionization-yield sideband; and $N_\text{DM}^\text{SB}$, the number of DM-search events in the ionization-yield sideband, via
\begin{align}
N_\text{DM}^\text{NR} =  N_\text{DM}^\text{SB} \left(N_\text{Ba}^\text{NR} \middle/ N_\text{Ba}^\text{SB} \right).
\end{align}

The model dataset for the \pbten{}-chain surface background takes advantage of events due to two low-activity ($\sim$0.1~Hz) \pbten{} sources installed in the Soudan cryostat directly adjacent to the surfaces of two of the detectors.  Although most of the exposure of these two detectors is used for the DM search, we left the first three months of data unblinded to make the surface rejection measurement cited previously~\cite{r133izip}, and we also utilize this unblinded data for the \pbten{}-chain model dataset.  Events from this dataset in the nuclear-recoil acceptance region are smeared to simulate noise differences among the detectors and reweighted based on relative detector efficiencies.  We assume their distributions would be otherwise identical between detectors. Because the \po{} $\alpha$ events can be unambiguously identified given their high energy (5.3~MeV), and the $^{210}$Po and $^{210}$Pb isotopes were in secular equilibrium during these datasets, we normalize the model dataset by the ratio of \po{} $\alpha$'s observed in the unblinded dataset to that observed in the relevant detector during the full DM-search dataset.


We use single-scatter \cf{} calibration data to model radiogenic- and cosmogenic-neutron backgrounds.  We reweight and normalize the calibration data to match the recoil energy spectra and event rate determined from Geant4~\cite{geant4:1} Monte Carlo simulations of these backgrounds, which predict 0.13 neutron events after pre\-selection.

\begin{figure}[t!!]
\begin{center}
\includegraphics[width=250pt]{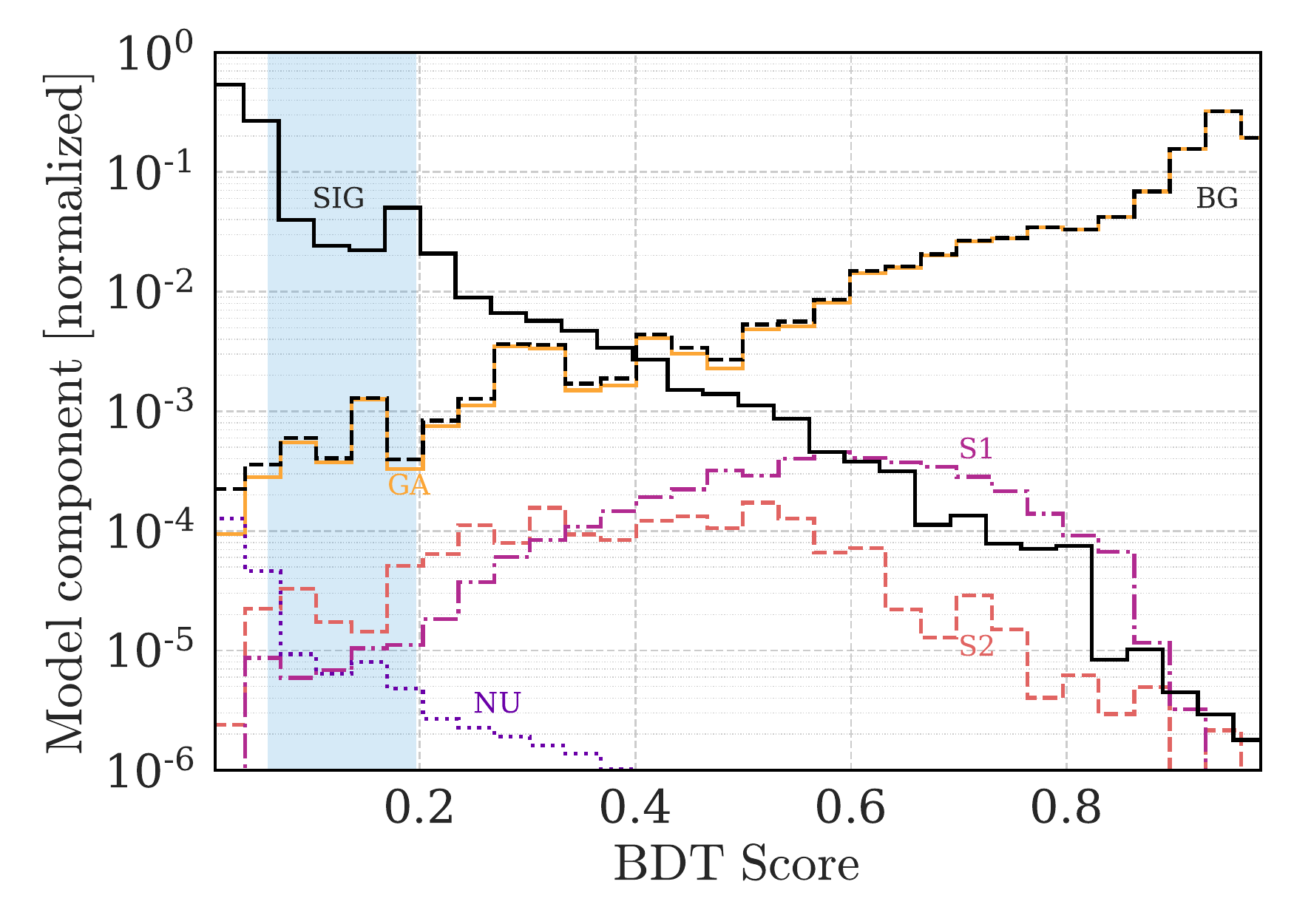}
\caption{ Histograms comparing background (BG: black dashed) and 50 GeV/$c^2$~signal models (SIG: black solid) in BDT score for the preselected events, summed over all detectors. The background model has been subdivided into its constituent components, which are, from darkest to lightest: radiogenic and cosmogenic neutrons (NU: purple dotted), upper surface \pbten{} chain (S1: fuchsia dot-dashed), lower surface \pbten{} chain (S2: coral dashed), and gammas (GA: orange). Both the signal and total-background model histograms have been normalized to unit integral for ease of comparison. The range of optimized BDT cut positions for the ten detectors used in this analysis is also shown (blue band).}
\label{fig:BDT}
\end{center}
\end{figure}

To define the DM acceptance region, we use a Gradient Boosted Decision Tree (BDT) approach~\cite{Friedman2001}.  It combines multiple input parameters to produce a single output parameter, the ``score,'' that quantifies how ``signal-like'' and ``background-like'' each event is, as shown in Figure~\ref{fig:BDT} for a 50 \gev~DM particle mass.   The input variables to the BDT are the recoil energy, ionization energy, ionization yield, and the phonon and ionization radial and $z$ partitions.  The partition quantities enable the BDT to optimize the fiducial volume accepted, while the energy quantities enable the BDT to use spectral differences to distinguish signal from backgrounds.  With ionization yield as an input, the BDT also further restricts the nuclear-recoil acceptance region.

We optimize the BDT-score selection as follows.  First, we find the combination of cut positions on the detectors' BDT scores that maximizes the total SAE for a particular DM particle mass, subject to the constraint that the total expected number of misidentified background events match a desired value.  The constraint is varied over the interval $[0,1)$.
  We then simulate 5,000 experiments for each mass/constraint pair. Events are sampled from our background models, and a 90\% C.L. upper limit on the DM-nucleon scattering cross section for each experiment is set by applying the optimum interval method without background subtraction~\cite{Yellin2002}.  For each candidate DM particle mass, a  BDT-score cut set---one cut for each detector---that approximately maximizes the average cross-section sensitivity over the simulated ensemble is identified. The set of cuts optimized for a 50 \gev~DM candidate is selected to define our final BDT-score selection because it has the best overall performance in the 10--250 \gev~ mass range.

\begin{figure}[t!!]
\begin{center}
\includegraphics[width=250pt]{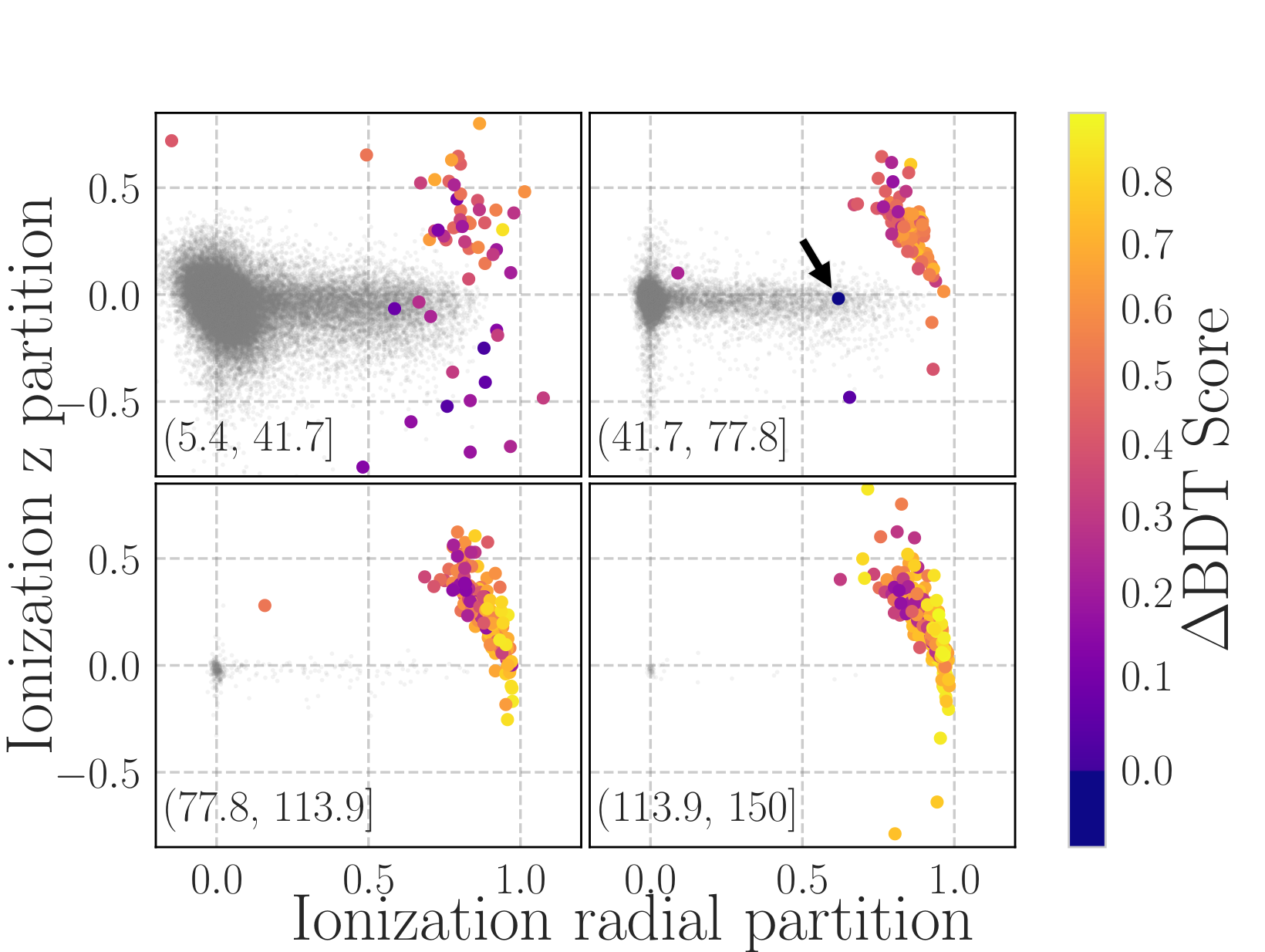}
\caption{ Scatter plots of ionization $z$ vs.\ radial partitions for all DM-search events passing preselection cuts (large, colored) and signal model events passing the preselection and BDT cuts (small, gray). The events are divided into four even energy bins, labeled in keV. The events for all ten detectors are present, and each DM-search event has been colored by the distance from the BDT cut position in the detector that registered the event to the BDT score of the event itself. This sets the BDT cut position at $\Delta$BDT $= 0$~and allows BDT scores to be compared between detectors. The single event accepted by the BDT cut is indicated with an arrow (and has $\Delta$BDT $< 0$).}
\label{fig:discrimination_plot}
\end{center}
\end{figure}

Unblinding the data after the final BDT cut reveals one DM candidate (42.8 keV recoil energy), as shown in Figure~\ref{fig:discrimination_plot}. This result is consistent with our model's expected misidentified-background distribution, which is approximately Poisson with a mean of 0.33 and predicts 1 ($\geq$1) background event in 24\% (28\%) of MC experiments.

\begin{figure}[b!!]
\begin{center}
\includegraphics[width=250pt]{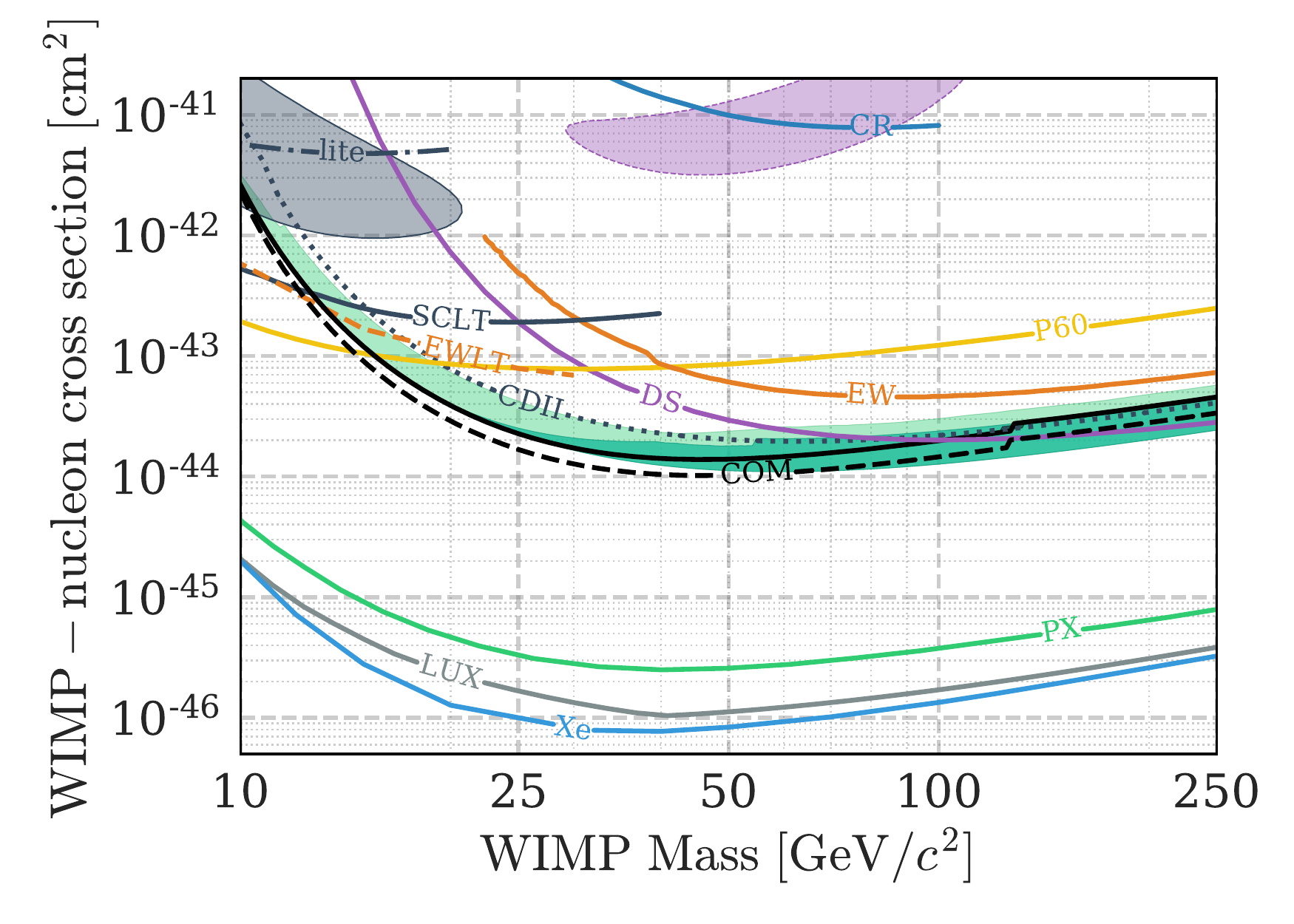}
\caption{ The $90\%$ confidence upper limit on the DM-nucleon cross section (solid black) based on our single observed event. The range of the pre-unblinding 68\% (95\%) most likely expected upper limits are shown as dark green (light green) bands. Closed contours shown are CDMS II Si~\cite{Agnese2013} (solid gray, 90\% C.L.) and DAMA/LIBRA~\cite{Savage2009} (dotted purple, 90\% C.L.). The remaining 90\% C.L. exclusion limits shown are, in order of increasing sensitivity at 25 GeV/$c^2$,  CRESST (CR)~\cite{Angloher2016}, CDMSlite Run 2 (lite)~\cite{r134cdmslite}, EDELWEISS (EW)~\cite{Armengaud2011329}, SuperCDMS Soudan low threshold (SCLT)~\cite{r133lowthreshold}, DarkSide (DS)~\cite{DarkSide}, PICO-60 (P60)~\cite{Amole2017}, EDELWEISS low mass (EWLT)~\cite{Armengaud2012}, CDMS II Ge alone (CDII)~\cite{c58GeRe} as well as a combined limit with this result (COM),  PandaX-II (PX)~\cite{PandaX}, LUX (LUX)~\cite{LUX2017}, and XENON1T (Xe)~\cite{Aprile:2017}.}

\label{fig:limit_plot}
\end{center}
\end{figure}

The optimal interval technique~\cite{Yellin2002} without background subtraction provides a 90\% C.L. upper limit on the DM-nucleon cross section, shown in Figure~\ref{fig:limit_plot}. The calculation uses the DM-particle and halo models summarized in~\cite{Lewin1996, Donato:1998pc}.  The resulting limit excludes new parameter space for DM--germanium-nucleus interactions in the mass range 13--127~\gev.  Using standard scalings~\cite{Lewin1996} between nuclei for spin-independent DM-nucleon interactions, limits obtained with other nuclei can be compared and are overlaid in Figure~\ref{fig:limit_plot}.

This work is the first analysis on the majority of the SuperCDMS Soudan dataset and is also the first analysis to fully utilize the background rejection power of the iZIP detector. By refining our background models and employing maximum likelihood techniques, future analyses may obtain improved sensitivity.

The SuperCDMS collaboration gratefully acknowledges technical assistance from the staff of the Soudan Underground Laboratory and the Minnesota Department of Natural Resources. The iZIP detectors were fabricated in the Stanford Nanofabrication Facility, which is a member of the National Nanofabrication Infrastructure Network, sponsored and supported by the NSF. Part of the research described in this article was conducted under the Ultra Sensitive Nuclear Measurements Initiative and under Contract No. DE-AC05-76RL01830 at Pacific Northwest National Laboratory, which is operated by Battelle for the U.S. Department of Energy. Funding and support were received from the National Science Foundation, the Department of Energy, Fermilab URA Visiting Scholar Award 15-S-33, NSERC Canada, and MultiDark (Spanish MINECO). Fermilab is operated by the Fermi Research Alliance, LLC under Contract No. De-AC02-07CH11359. SLAC is operated under Contract No. DEAC02-76SF00515 with the United States Department of Energy.

\bibliography{SuperCDMS_HT}

\end{document}